\documentclass[aps,apl,preprint]{revtex4}
\usepackage{graphicx}
\begin{document}

\title{Determination of spin polarization in InAs/GaAs self-assembled quantum dots}

\author{F. G. G. Hernandez}
\email{felix@ifi.unicamp.br} \affiliation{Laborat\'orio Nacional
de Luz S\'{\i}ncrotron, Caixa Postal 6192 - CEP 13084-971,
Campinas, SP , Brazil} \affiliation{Instituto de F\'isica Gleb
Wataghin, Universidade Estadual de Campinas, Campinas, SP ,
Brazil}
\author{T. P. Mayer Alegre}
\affiliation{Laborat\'orio Nacional de Luz S\'{\i}ncrotron, Caixa
Postal 6192 - CEP 13084-971, Campinas, SP , Brazil}
\affiliation{Instituto de F\'isica Gleb Wataghin, Universidade
Estadual de Campinas, Campinas, SP , Brazil}
\author{G. Medeiros-Ribeiro}
\email{gmedeiros@lnls.br} \affiliation{Laborat\'orio Nacional de
Luz S\'{\i}ncrotron, Caixa Postal 6192 - CEP 13084-971, Campinas,
SP , Brazil}

\date{\today}

\begin{abstract}
The spin polarization of electrons trapped in InAs self-assembled
quantum dot ensembles is investigated. A statistical approach for the
population of the spin levels allows one to infer the spin
polarization from the measure values of the addition energies. From the
magneto-capacitance spectroscopy data, the authors found a fully polarized
ensemble of electronic spins above 10 T when $\mathbf{B}\parallel[001]$ and at 2.8 K.
Finally, by including the g-tensor anisotropy the angular
dependence of spin polarization with the magnetic field
$\mathbf{B}$ orientation and strength could be determined.
\end{abstract}

\maketitle

Spin polarization in quantum dots (QDs) is a desirable measurement for a complete
characterization of the magnetic response of these systems. The assessment of this quantity
has an important impact on the usefulness of QDs for quantum computing
processing schemes. Recently, there has been several experiments relying on the
optical selection rules for the determination of spin polarization in QD ensembles \cite{gupta,
chye,loffler,li,lombez,gundogdu,itskos}.
State preparation and measurement with circularly polarized excitation
\cite{gupta}, spin injection with ferromagnetic contacts into light emitting diodes \cite{chye,loffler,li,lombez}, time and polarization resolved photoluminescence \cite{gundogdu} and oblique Hanle effect
\cite{itskos} have been used in order to characterize spin polarization. Yet, in order to explain the inferred polarization, theoretical results
point out that one needs to take into account the hole polarization in the final result \cite{pryor,hawrylak}.

In contrast to optical schemes, electrical readout of the
electronic spins orientation does not require any knowledge of hole
polarization, depending only on spin-to-charge conversion \cite{vandersypen}.
Here we perform magneto-capacitance measurements of
QDs embedded in a Metal-Insulator-Semiconductor (MIS) capacitor
structure. The amount of polarization was inferred by measuring
the electron addition energies, from which the Helmholtz free energy associated
to the spin degree of freedom was extracted.

InAs QDs capped with thin InGaAs strain reducing layers were grown
by molecular beam epitaxy at a temperature of 530 $^\circ$C as
described elsewhere \cite{gilbertoapl,gilbertogfactor}. Schottky
diodes were subsequently defined by conventional photolithography.
The area of the devices was 4$\times$$10^{-4}$ cm$^{2}$,
encompassing about 10$^{7}$ QDs per diode. The capacitance measurements were
performed at a nominal temperature of 2.8K using lock-in
amplifiers at a frequency of 7.5KHz. An ac amplitude of 4 mV(rms)
was superimposed on a varying dc bias ranging from -2 V to 0.5 V.
The experiments were carried out in a superconducting magnet
for intensities ranging from 0 to 15 T. The orientation of the magnetic field with respect
to the sample crystallographic axes could be adjusted with a
goniometer with a precision better 0.5$^\circ$.

The gate bias and the chemical potential inside the QDs can be related by solving the Poisson equation
for the MIS structure \cite{gilbertogfactor}. In a capacitance-voltage measurement, electrons are
sequentially loaded into QDs at selected biases and thus the addition energies can be
inferred by evaluating the chemical potential $\mu$ for each
electronic configuration. The addition energies indicate how much
energy is required to add the {\it i}-th electron compared to the
energy needed to add the ({\it i}-1)th electron \cite{he}, namely
$\Delta\mu=\mu_{i}-\mu_{i-1}$.

The InAs QDs electronic structure can be described by a lateral parabolic confining potential
\cite{warburton,warburton98,raymond}. Furthermore, the observation of the
Aufbau principle and Hund's rule in the charging process confirm
that the Fock-Darwin parabolic approximation gives a simple
yet precise description for the energy ladder in InAs/GaAs
QDs \cite{he}.

In addition to the quantum confinement contribution, the chemical
potential includes the magnetic field dependent Coulomb
interaction E$_{C}$(B) \cite{prlthiago,formula}, and the Helmholtz
free energy $F_{J}(B,T)$ that accounts for the spin contribution,
where $J$ is the total angular momentum. Defining $S=\sum S_{z}$,
$ L= \sum L_{z}=0 $ for the s-shell, and $J=L+S$, the values
corresponding to the charging of one and two electrons are
$J=1/2$, $J=0$. $F_{J}(B,T)$ can be calculated as:

\begin{equation}
F_{J}(B,T)=-kTlnZ_{J}(B,T)
\end{equation}
where $Z_{J}(B,T)=\sum_{m_{J}=-J}^{m_{J}=J}e^{x}$ is the partition
function of the system with $x=m_{J}g\beta B/k_{B}T$ denoting the
relation between the magnetic and thermal energy, where $\beta$ is
the Bohr magneton, B is the magnetic field strength, and {\it g}
is the orientation dependent Land\'{e} g-factor described as
$g=\sqrt{g^{2}_{[001]}cos^{2}(\theta)+g^{2}_{[110]}sin^{2}(\theta)}$,
where $\theta$ is the angle between {\bf B} and the [001]
direction \cite{prlthiago,abragam}.

The spin polarization $M$ can be defined as the difference between
the populations of the up and down spin levels normalized by total
number of spins. The interaction between neighboring dots is negligible, and thus
will not be considered here. Due to the strong confinement, the spin splitting
observed in QDs is mostly due to the Zeeman effect, and therefore the Rashba and Dresselhaus
contributions are minimal \cite{hanson}.

The derivative of the Helmholtz free energy gives a relation
for the number of magnetic moments aligned with $\mathbf{B}$ at a
constant temperature, from which the polarization can be expressed:
\begin{equation}
M=\partial F_{J}(B,T)/\partial B
\end{equation}

The normalized polarization is defined by the Brillouin functions.
For $J=1/2$, $M=1/2g\beta B_{1/2}(B,T)$ where
$B_{1/2}(B,T)=tanh(g\beta B/2k_{B}T)$. If the spins are perfectly
polarized, the magnetic field splits the spin levels linearly with
slopes $\pm 1/2g\beta B$. Thus, a constant derivative of
F$_{J}$(B,T) when changing the magnetic field implies in a fully
polarized spin system.

Expressing the addition energies including electrostatic,
quantum confinement, and spin terms we obtain for the first 2
particle levels \cite{gilberto97}:

\begin{eqnarray}
E_{1}&=&E_{z}+\hbar\Omega-kTln\left[2cosh\left(g\beta B/2kT\right)\right]-\mu_{1} \\
E_{2}&=&2E_{z}+2\hbar\Omega+E_{C}(B)-2\mu_{2} \nonumber
\end{eqnarray}
where $E_{z}$ is the confining energy in the growth direction
and $\Omega$ is the {\bf B} dependent natural frequency of the
harmonic confining potential, defined by
$\sqrt{\omega^{2}_{0}+\omega^{2}_{c}/4}$ where $\omega_c$ is the cyclotron frequency
and $\omega_0$ is the natural frequency associated with the
lateral parabolic confinement. $\mu_{1}$ and $\mu_{2}$
can be calculated by equating $E_1=0$ and $E_2=E_1$, respectively:

\begin{eqnarray}
\mu_{1}&=&E_{z}+\hbar\Omega-kTln\left[2cosh\left(g\beta B/2kT\right)\right] \\
\mu_{2}&=&E_{z}+\hbar\Omega+kTln\left[2cosh\left(g\beta B/2kT\right)\right]+E_{C}(B)
\nonumber
\end{eqnarray}

These leads to the addition energy for the s shell:

\begin{equation}\label{diferences}
\Delta\mu=2kTln[2cosh(g\beta B/2kT)]+E_{C}(B)
\end{equation}
which, in the limit of large $g\beta B/kT$, gives the expected
Zeeman splitting $g\beta B$.

\begin{figure}
\includegraphics{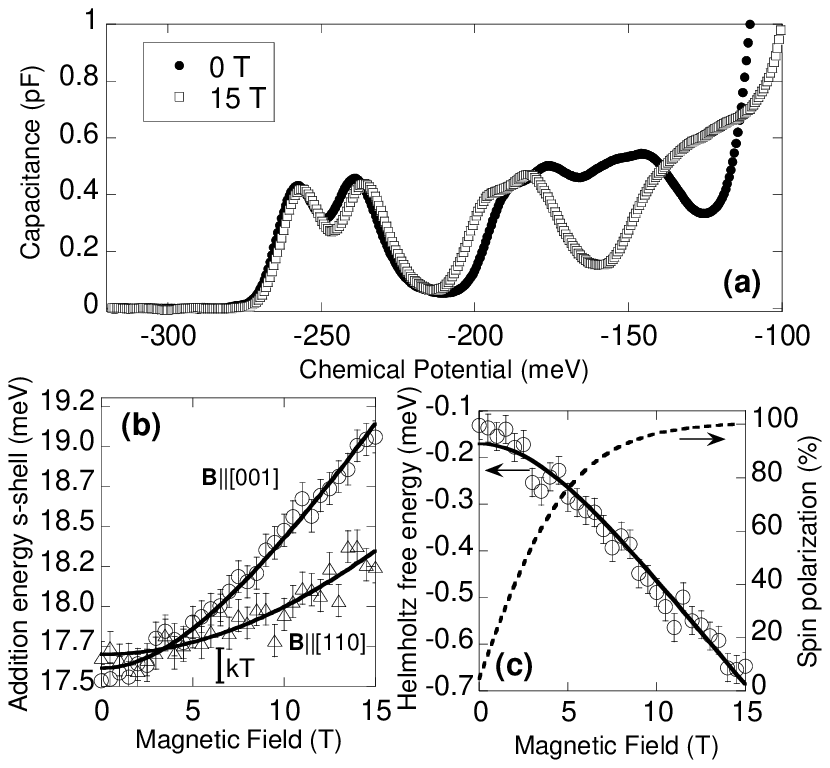}%
\caption{(a) Capacitance spectra taken at zero (solid circles) and
15 T (open squares) for $\mathbf{B}\parallel[001]$. The peaks
associated to the s and p states can be easily identified, as well
as the effect of the magnetic field on the orbital properties. The
background corresponding to the geometrical capacitance has been
removed. (b) Addition energy for n=2 (s-shell filling) as a
function of the applied magnetic field for
$\mathbf{B}\parallel[001]$ (open circles) and
$\mathbf{B}\parallel[110]$ (open squares). The solid lines are
fits using equation \ref{diferences}. (c) Helmholtz Free Energy
$F_{1/2}$ (solid lines) and the experimental data, extracted from
the $\mu_1$ dependence on the magnetic field, where we retained
only the spin contribution; the orbital term $\hbar\Omega$ and the
z confinement energy $E_z$ were subtracted. Spin polarization
(dashed line) associated with the derivative of the measured
addition energy plotted analytically as $tanh(g\beta B/2k_BT)$
using the inferred g factor and temperature. \label{f1}}
\end{figure}

Capacitance spectra at two different magnetic fields are shown in
figure 1a. In order to analyze the data from an ensemble of
quantum dots, we fit gaussian functions to each tunneling event.
Solving the sum and difference between the position of the s-state
gaussian peaks, we obtain $\hbar\omega_{0}=37.8\pm0.2meV$,
$E_{z}=280meV\pm2meV$, $E_{C}(0)=17.2meV\pm0.2meV$. The
electrostatic interaction $E_C$ can be calculated from
$\hbar\omega_0$ in the parabolic confinement assumption
\cite{warburton}, with an agreement better than 5-10\% with the
measured values of $E_C$.

In our analysis, we fit the peak difference using the equation
\ref{diferences} as showed in figure 1b, using the g-factor as the
fitting parameter. The g factor for the s shell is found to be
anisotropic \cite{prlthiago}: $g_{[001]}=1.51 \pm 0.05$ and
$g_{[110]}=0.87 \pm 0.05$. For $B\rightarrow0$, $\Delta\mu\sim
kT$, which can be verified in figure 1b.

The main feature in figure 1b is that the addition energy
evolution with the magnetic field is not linear for fields below
approximately 5 T. This is directly related with the lack of spin
orientation for the loaded electrons due to thermal disorder
acting to randomize the magnetic moments. In figure 1c we plot the
Helmholtz free energy $F_{J}(B,T)=-kTln[2cosh(g\beta B/2kT)]$,
extracted from the position $\mu_{1}$ of the first capacitance
peak as a function of B. The polarization M is shown by the dashed
line, calculated from the functional form of $F_{J}(B,T)$ and the
fitting parameter g. We can identify 5 T as the field at which
80\% of all spins are polarized with $\mathbf{B}||[001]$. For
$\mathbf{B}$ parallel to the plane, this point is higher due to a
smaller g-factor. The inferred polarization agrees with the
empirical value utilized in an experiment of spin-selective
optical absorption at 8 T \cite{90}.

\begin{figure}
\includegraphics{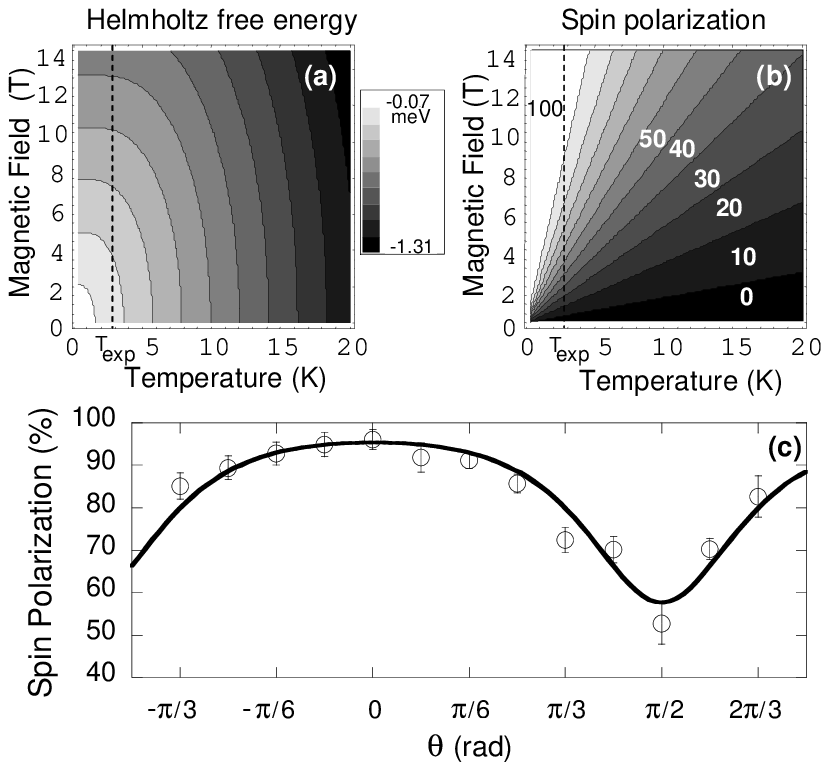}
\caption{ (a) Calculated Helmholtz Free Energy $F_{J}$ as a
function of the magnetic field and temperature, evaluated from the
measured values of g for the s shell for $\mathbf{B}||[001]$. (b)
Associated spin polarization $\partial F_{J}/\partial B$. The
temperature utilized during the experiments is represented by the
dashed lines. (c) Angular dependence of the spin polarization for
B=10T and T=2.7K. The experimental data is shown by empty circles.
\label{f2}}
\end{figure}

Finally, we can infer from these results the polarization
dependence on the temperature. Figure 2a shows that the Helmholtz
free energy dependence on B and T. Above 15 K, the spin
contribution to the addition energy is negligible. Figure 2b shows
that for 5 K, field in excess of 15 T are required to fully
polarize the spins. In figure 2c, the angular dependence of the
polarization is shown. The degree of polarization varies between
90 and 50\%, in agreement with a recent theoretical investigation
\cite{hawrylak}.

In summary, we were able to determine the degree of polarization
of a system of non-interacting spins at finite temperatures and
magnetic fields. The angular dependence reflects a spin
polarization anisotropy due to the g-tensor characteristics, which
should be taken into account in the interpretation of polarization
experiments.

This work was supported by the MCT-CNPq and HP-Brazil. FGGH and
TPMA would like to acknowledge FAPESP contracts 04/02814-6 and
04/01228-6 for financial support. We thank the IFGW-UNICAMP for
the use of the high magnetic field facility.

\end{document}